\documentclass[prd,tightenlines,floatfix,showpacs,preprintnumbers,nofootinbib,eqsecnum]{revtex4}

 \usepackage[dvips,final]{graphicx}
  \usepackage{amssymb}
   \usepackage{amsmath} 
    \usepackage{amsfonts}
     \usepackage{epsfig}
     \usepackage{color}
      \usepackage{bm} 
        \usepackage{xcolor}
        
\usepackage{booktabs}
\usepackage{multirow}
\usepackage{slashed}
\usepackage{comment}

\def\doublespace{\def\baselinestretch{1.6}\large\normalsize}
\def\normalspace{\def\baselinestretch{1.0}\normalsize}
\def\PSfig#1#2{\scalebox{#1}{\includegraphics{#2}}}

\def\Caption#1{
  \normalspace
  \vskip-1mm\caption{\sl#1}\vskip-1mm
  \doublespace
}

\def\BA{\begin{eqnarray}}
\def\BE{\begin{equation}}
\def\BF{\begin{figure}[htb]}
\def\BT{\begin{table}[htb]}
\def\EA{\end{eqnarray}}
\def\EE{\end{equation}}
\def\EF{\end{figure}}
\def\ET{\end{table}}

\def\la{\langle}
\def\ra{\rangle}
\def\fm{\,\mbox{fm}}

\def\mb{\,\mbox{mb}}

\def\GeV{\,\mbox{GeV}}

\def\Jpsi{J\!/\!\psi}
\def\psip{\psi^{\,\prime}}

\def\Y{\Upsilon}
\def\Yp{\Upsilon^{\,\prime}}

\def\sqq{\sigma_{Q\bar Q}}

\def\aem{\alpha_{em}}

\def\lsim{\mathrel{\rlap{\lower4pt\hbox{\hskip1pt$\sim$}}
     \raise1pt\hbox{$<$}}}         
 \def\gsim{\mathrel{\rlap{\lower4pt\hbox{\hskip1pt$\sim$}}
     \raise1pt\hbox{$>$}}}         

\begin{document}

\title{
Electroproduction of heavy quarkonia: significance of dipole orientation
}

\author{B. Z. Kopeliovich$^1$}
\email{boris.kopeliovich@usm.cl}

\author{M. Krelina$^{2,3}$}
\email{michal.krelina@cvut.cz}

\author{J. Nemchik$^{2,4}$}
\email{nemcik@saske.sk}

\affiliation{$^1$
Departamento de F\'{\i}sica,
Universidad T\'ecnica Federico Santa Mar\'{\i}a,
Avenida Espa\~na 1680, Valpara\'iso, Chile}
\affiliation{$^2$
FNSPE, Czech Technical University in Prague, B\v rehov\'a 7, 11519
Prague, Czech Republic}
\affiliation{$^3$
Physikalisches Institut, University of Heidelberg, Im Neuenheimer Feld 226, 69120 Heidelberg, Germany}
\affiliation{$^4$
Institute of Experimental Physics SAS, Watsonova 47, 04001 Ko\v sice, Slovakia
}

\date{\today}
\begin{abstract}
\vspace*{5mm}

The differential cross section $d\sigma/dq^2$ of diffractive electroproduction of heavy quarkonia on protons is a sensitive study tool for the interaction dynamics within the dipole representation. 
Knowledge of the transverse momentum transfer $\vec q$ provides a unique opportunity to identify the reaction plane, due to a strong correlation between the directions of $\vec q$ and impact parameter $\vec b$. On top of that, the elastic dipole-proton amplitude is subject to a strong correlation between $\vec b$ and dipole orientation $\vec r$. Most of models for $b$-dependent dipole cross section either completely miss this information, or make unjustified assumptions. We perform calculations basing on a realistic model for $\vec r$-$\vec b$ correlation, which significantly affect the $q$-dependence of the cross section, in particular
the ratio of $\psip(2S)$ to $J/\psi$ yields. We rely on realistic potential models for the heavy quarkonium wave function, and the Lorentz-boosted Schr\"odinger equation.
Good agreement with data on $q$-dependent diffractive electroproduction of heavy quarkonia  is achieved.

\end{abstract}


\maketitle

%
%
%
\section{Introduction}
\label{intro}
%
%
%

Elastic real and virtual photoproduction of heavy quarkonia on protons is an effective tool for study of the space-time pattern of diffraction mechanism, as well as related aspects of quantum-chromodynamics (QCD). 
The long-standing history of our investigation of the photo- and electroproduction of vector mesons 
\cite{Kopeliovich:1991pu,Kopeliovich:1993gk,Kopeliovich:1993pw,Nemchik:1994fq,Nemchik:1994fp,Nemchik:1996pp,Nemchik:1996cw,Kopeliovich:1999am,jan-00a,jan-00b,Kopeliovich:2001xj,Nemchik:2002ug,Kopeliovich:2007wx,Krelina:2018hmt,Cepila:2019skb,Krelina:2019egg,Kopeliovich:2020has,Krelina:2020bxt}
has provided the foundations for theoretical interpretation and
has contributed in an essential way to understanding of this process within the QCD color dipole formalism. Such a formalism has been  frequently used recently in the literature with only minor improvements in corresponding model descriptions.

The most of experimental data are available only for the cross section integrated over transfer momenta.
However, an additional information about the dynamics and
properties of the diffraction mechanism itself
can be acquired by studying also the momentum transfer dependence of real and virtual photoproduction cross sections. Knowledge of the the transverse momentum transfer $\vec q$ provides a unique opportunity to identify the reaction plane, because the Fourier transformation from impact parameters $\vec b$ to momentum representation leads to a strong correlation between $\vec b$ and $\vec q$.

Within the light-front (LF) color dipole formalism, the amplitude for electroproduction of heavy vector mesons with the transverse momentum transfer $\vec q$ can be expressed in the factorized form,
%
\BA
\mathcal{A}^{\gamma^\ast p\to V p}(x,Q^2,\vec q)
=
\bigl\la V |\tilde{\mathcal{A}} |\gamma^*\bigr\ra
=
\int d^2r\int_0^1 d\alpha\,
\Psi_{V}^{*}(\vec r,\alpha)\,
\mathcal{A}_{Q\bar Q}(\vec r, x, \alpha,\vec q)\,
\Psi_{\gamma^\ast}(\vec r,\alpha,Q^2)\,
\label{amp-p0}
\EA
%
using thus the advantage of the $(\vec r,\alpha)$ diagonalization of the scattering matrix $\tilde{\mathcal{A}}$ with the normalization $(d\sigma/dq^2)_{q=0} = |\mathcal{A}|^2/16\,\pi$.
Here
$\mathcal{A}_{Q\bar Q}(\vec r, x, \alpha,\vec q)$ is the amplitude for elastic scattering of the color dipole
on the nucleon target,
$\Psi_V(r,\alpha)$ is the LF wave function for heavy quarkonium and
$\Psi_{\gamma^\ast}(r,\alpha,Q^2)$ is the LF distribution 
of the $Q\bar Q$ Fock component of the real ($Q^2 = 0$) or virtual ($Q^2 > 0$) photon,
where $Q^2$ is the photon virtuality and the $Q\bar Q$ fluctuation (dipole) has the transverse size $\vec{r}$.
The variable $\alpha$ is the fractional LF momentum carried by a heavy quark or antiquark from a $Q\bar Q$ Fock component of the photon and $x =  (m_V^2 + Q^2 -t)/(W^2 + Q^2) = (m_V^2+Q^2-t)/s$, where $m_V$ is the quarkonium mass, $W$ is c.m. energy of the photon-nucleon system and $t = -q^2$.

In Eq.~(\ref{amp-p0}) the amplitude $\mathcal{A}_{Q\bar Q}(\vec r, x, \alpha,\vec q\,)$ can be written in terms of the gluon density matrix $\mathcal{F}(x,\vec{k},\vec{k}^{\,\prime},\vec q\,)$, which is proportional to the imaginary part of the non-forward gluon-nucleon scattering amplitude. In the limit of $\vec q \to 0$ one arrives at the unintegrated gluon structure function of the nucleon, $\mathcal{F}(x,\vec{k},\vec{k}^{\,\prime},\vec q=0) = \mathcal{F}(x,\vec{k},\vec{k})\equiv \mathcal{F}(x,k^2)= \partial\,G(x,k^2)/\partial\,log\,k^2$, and the general formula for $\mathcal{A}_{Q\bar Q}(\vec r, x, \alpha,\vec q)$ at $\vec q = 0$ should access the standard expression for the dipole cross section, 
%
\BA
\sqq(r,x)
=
\frac{4\pi}{3}\int\frac{d^{2}k}{k^{4}}\,
\Bigl [1-e^{-i\vec{k}.\vec{r}}\Bigr ]\,
\alpha_{s}(k^{2})\mathcal{F}(x,k^2)
=
\frac{\pi^2\,r^2}{3}
\int\frac{dk^2}{k^{2}}\,
\frac{4\,\bigl [1-J_0(k\,r)\bigr ]}{(k\,r)^2}\,
\alpha_{S}(k^{2})\mathcal{F}(x,k^2)\,,
\label{dcs-k}
\EA
%
which leads for small dipole sizes $r\ll r_0$, where $r_0\sim 0.3\,\fm$ represents the gluon propagation radius \cite{spots,drops}), to the following well known expression,
%
\BA\sqq(r,x)
=
\frac{\pi^2}{3}\,r^2\,\alpha_S(r)\,G(x,k_s^2)\,,
\label{dcs-r2}
\EA
%
where the gluon structure function is scanned at the factorization scale $k_s^2\sim A_s/r^2$, with the large factor $A_s\sim 9\div 10$ determined in Refs.~\cite{Nikolaev:1993th,Nikolaev:1994ce}.

The most of phenomenological studies of momentum transfer dependence of differential cross sections are usually performed within the familiar impact parameter representation for the color dipole elastic scattering amplitude $\mathcal{A}_{Q\bar Q}(\vec r, x, \alpha,\vec b)$ related to $\vec q$-dependent amplitude via Fourier transform,
%
\BA
\mathcal{A}_{Q\bar Q}(\vec r, x, \alpha,\vec q)
=
\int d^2 b\,e^{- i \vec b\cdot\vec q}\,
\mathcal{A}_{Q\bar Q}(\vec r, x, \alpha,\vec b)
\EA
%
with the correct reproduction of the dipole cross section at $\vec q=0$,
%
\BA
\sqq(r,x)
=
\mathrm{Im}\mathcal{A}_{Q\bar Q}(\vec r, x, \alpha,\vec q=0) 
= 
2\,\int d^2 b\,
\mathrm{Im}\mathcal{A}^N_{Q\bar Q}(\vec r, x, \alpha,\vec b)\,,
\label{dcs-b}
\EA
%
where 
the \textit{partial} dipole elastic amplitude
$\mathrm{Im}\mathcal{A}^N_{Q\bar Q}(\vec r, x, \alpha,\vec b)$
represents the interaction of the $Q\bar Q$ dipole
with a nucleon target at impact parameter $\vec b$. 

Consequently, the next very important step is to determine the partial amplitude $\mathrm{Im}\mathcal{A}^N_{Q\bar Q}(\vec r, x, \alpha,\vec b)$ 
including a proper
correlation between the color dipole orientation $\vec r$ and the impact parameter of a collisions $\vec b$. 
There are several widely used models for the $b$-dependent dipole cross sections with additional $b$-dependent part giving so the partial dipole amplitude without 
an adequate $\vec b$-$\vec r$ correlation, i.e. 
assuming usually that the scattering amplitude is independent of the angle between the vectors $\vec b$ and $\vec r$
(see  Ref.~\cite{Rezaeian:2012ji} for b-IPsat model, Ref.~\cite{Kowalski:2003hm} for b-Sat model,
Ref.~\cite{Kowalski:2006hc,Rezaeian:2013tka} for b-CGC model, Ref.~\cite{Cepila:2018faq} for b-BK model, for example).

In the present paper we investigate for the first time the momentum transfer dependence of differential cross sections in elastic photo- and electroproduction of heavy quarkonia on protons
using a proper color dipole orientation without any approximation within standard phenomenological models for dipole cross sections of the saturated form. The corresponding partial dipole elastic amplitude is based on our previous studies \cite{Kopeliovich:2008dy,Kopeliovich:2007sd,Kopeliovich:2007fv,Kopeliovich:2008nx} and includes a proper correlation between vectors 
$\vec b$ and $\vec r$ as is described in Sec.~\ref{amp-born}. Here we shortly illustrate how the color dipole orientation looks like within the simplified Born approximation. 
In the next Section~\ref{sat-cdm},
we present the explicit form for the partial $Q\bar Q$-proton 
amplitude with parameters corresponding to GBW
\cite{GolecBiernat:1998js,GolecBiernat:1999qd} and
BGBK
\cite{Bartels:2002cj}
saturation models for the dipole cross section.
Section~\ref{vm-proton} contains expressions for calculation of $t$-dependent differential cross sections as well as forward diffraction slopes within the LF color dipole formalism. In order to exclude a spurious $D$-wave admixture, here we treat a simple non-photon-like structure of the $V\to Q\bar Q$ transition in the $Q\bar Q$ rest frame as in our previous studies \cite{Krelina:2018hmt,Cepila:2019skb,Krelina:2019egg,Kopeliovich:2020has,Krelina:2020bxt}. This requires to perform, besides the standard Lorentz boost \cite{Terentev:1976jk} to the LF frame of radial components of quarkonium wave functions,
also transformation of the corresponding spin-dependent parts known as \textit{the Melosh spin rotation}
\cite{Melosh:1974cu}. 
The next Sec.~\ref{data-proton} is devoted to comparison of our results for $d\sigma^{\gamma^\ast p\to \Jpsi p}/dt$ with available data from H1 and ZEUS experiments at HERA. Here we also present predictions for various quarkonium states, as well as for the $\psip(2S)$-to-$\Jpsi(1S)$ and $\Yp(2S)$-to-$\Y(1S)$ ratios $R_{V'/V}(t) = \{d\sigma^{\gamma^\ast p\to V' p}/dt\} / \{d\sigma^{\gamma^\ast p\to V p}/dt\}$
that can be confirmed by future measurements. 
Finally, the last Sec.~\ref{final} contains a summary with the main concluding remarks.

%
%
%
\section{Partial dipole amplitude in Born approximation}
\label{amp-born}
%
%
%

It was shown that azimuthal asymmetry of pions 
\cite{Kopeliovich:2008nx},
as well as direct photons \cite{Kopeliovich:2008dy,Kopeliovich:2007sd,Kopeliovich:2007fv} 
in $pp$ and $pA$ collisions is based on
the correlation between the color dipole orientation and the impact parameter of a
collision. 
In the present paper we analyse the impact of such correlation on 
the momentum transfer dependence of
differential cross sections of heavy quarkonium electroproduction on protons, $\gamma^\ast p\to V p$.

For a colorless heavy quark $Q\bar Q$ photon fluctuation (dipole) of transverse separation $\vec r$ with the impact parameter $\vec{b}$ of its center of gravity, the corresponding interaction of the $Q\bar Q$ dipole can occur due to difference between impact parameters of $Q$ and $\bar Q$ relative to the scattering center. 
This eliminates the production of any $Q\bar Q$ photon component with the same impact parameter
from the target related to $Q$ and $\bar Q$ independently of the magnitude of $\vec r$. Thus the dipole interaction vanishes if $\vec{r}\perp\vec{b}$ and is maximal if $\vec{r}$$\parallel$$\vec{b}$. Such a correlation between the vectors $\vec r$ and $\vec b$ can be seen in terms of the dipole partial elastic amplitude $\mathcal{A}_{Q\bar Q}^q(\vec r,\vec b)$ describing the dipole interaction with a quark in Born approximation (two-gluon exchange model),
%
 \BA
\mathrm{Im} \mathcal{A}_{Q\bar Q}^q(\vec r,\vec{b}) 
&=& 
\frac{\mathcal{N}}{(2\pi)^2} \int
\frac{d^2k\,d^2k^{\,\prime}}{(k^2+m_g^2)(k^{\,\prime\,2}+m_g^2)}\,
\left[e^{i\vec k\cdot(\vec b +\vec r/2)}- e^{i\vec k\cdot(\vec b -\vec
r/2)}\right]
\left[e^{i\vec k^{\,\prime}\cdot(\vec b +\vec r/2)}- e^{i\vec
k^{\,\prime}\cdot(\vec b -\vec r/2)}\right],\nonumber\\
&=& \mathcal{N}
\left[K_0\left(m_g\left|\vec b +\frac{\vec r}{2}\right|\right) -
K_0\left(m_g\left|\vec b -\frac{\vec r}{2}\right|\right)\right]^2,
\label{born}
 \EA
%
where $m_g$ represents an effective gluon mass accounting for the confinement and other nonperturbative effects;
$K_0(x)$ is the modified Bessel function of the second kind.
Here, for simplicity, we assume the same longitudinal momenta of $Q$ and $\bar Q$. 
The above Eq.~(\ref{born}) clearly demonstrates a correlation between $\vec r$ and
$\vec b $, as well as a vanishing of $\mathcal{A}_{Q\bar Q}^q(\vec r,\vec b)$ when $\vec b \cdot\vec r=0$.

The Born approximation for the dipole partial amplitude is rather crude since does not lead to energy dependent dipole cross section $\sqq(r,x)$.  For this reason,  it is worth switching to a more reliable model for $\mathcal{A}_{Q\bar Q}^q(\vec r,\vec b)$ without any approximation.
However, another alternative way how to include the energy dependence is based on an improvement of the Born approximation with corresponding modification of the coefficient $\mathcal{N}\Rightarrow \mathcal{N}(x)$
in Eq.~(\ref{born}) as is described in Ref.~\cite{Kopeliovich:2008nx}.

%
%
%
\section{Partial dipole amplitude in the saturation model}
\label{sat-cdm}
%
%
%

The partial dipole elastic amplitude has been introduced in Ref.~\cite{Kopeliovich:2007fv} within the standard model for the dipole cross section $\sqq$ of a conventional saturated form corresponding to
various phenomenological parametrizations proposed in the literature (see \cite{GolecBiernat:1998js,Kowalski:2006hc,Kopeliovich:1999am,Bartels:2002cj,Rezaeian:2012ji}, for example). They are based on the fits to the HERA DIS data and
thus includes contributions from higher order perturbative corrections, as well as nonperturbative effects. For a $Q\bar Q$ dipole interacting with a proton at impact parameter $\vec b$, such dipole 
amplitude $\mathcal{A}^{N}_{Q\bar{Q}}$ 
reads
\cite{Kopeliovich:2007fv},
%
 \BA
\mathrm{Im} \mathcal{A}^N_{Q\bar Q}(\vec r, x, \alpha,\vec b)
&=&
\frac{1}{12\pi}\,
\int\frac{d^2k\,d^2k^{\,\prime}}{k^2\,k^{\,\prime\,2}}\,
\sqrt{\alpha_s(k^2)\,\alpha_s(k^{\,\prime\,2})}\,\,
{\cal F}(x,\vec k,\vec k^{\,\prime})\,
e^{i\,\vec b \cdot(\vec k-\vec k^{\,\prime})}\,
\nonumber\\
&\times&
\left(e^{-i\,\vec k\cdot\vec r\alpha}-
e^{i\,\vec k\cdot\vec r\,(1-\alpha)}\right)\,
\left(e^{i\,\vec k^{\,\prime}\cdot\vec r\,\alpha}-
e^{-i\,\vec k^{\,\prime}\cdot\vec r\,(1-\alpha)}\right)\,
\,, 
\label{sat}
 \EA
%
 where $\mathcal{F}(x,\vec{k},\vec{k}^{\,\prime})$ is the generalized
 unintegrated gluon density. In the two-gluon exchange model, Eq.~(\ref{born}),  we assumed that $\alpha=1/2$ and thus
$Q$ and $\bar{Q}$ are equally distant from the dipole center of gravity.

The shape of 
$\mathcal{F}(x,\vec{k},\vec{k}^{\,\,\prime})$ in Eq.~(\ref{sat}) has been determined by comparing with the saturated form of the dipole cross section,
%
\BA
\sqq(r,x) =
\sigma_0\,
\biggl (1 - \exp \Bigl [ - \frac{r^2}{R_0^2(x)}\Bigr ] \biggr )\,,
\label{gbw}
\EA
%
using expression Eq.~(\ref{dcs-k}).
This gives
the off-diagonal unintegrated gluon density of the form~\cite{Kopeliovich:2007fv},
%
\BA 
{\cal F}(x,\vec k,\vec k^{\,\prime})
=
\frac{3\,\sigma_0}{16\,\pi^2\,\sqrt{\alpha_s(k^2)\,\alpha_s(k^{\,\prime\,2})}}\,\, k^2\,k^{\,\prime\,2}\,R_0^2(x)\,
{\rm exp}\Bigl[-\frac{1}{8}\,R_0^2(x)\,(k^2+k^{\,\prime\,2})\Bigr]\,
{\rm exp}\Bigl[- \frac{1}{2}\,R^2_N(x)(\vec k-\vec k^{\,\prime})^2\Bigr]\,,
\label{UGD} 
\EA
%
with a correct relation to the diagonal gluon density as $\mathcal{F}(x,\vec{k},\vec{k}^{\,\,\prime}=\vec{k})=\mathcal{F}(x,k^2)$.
Adopting  the GBW dipole model \cite{GolecBiernat:1998js,GolecBiernat:1999qd}, for example, the above parameters in Eq.~(\ref{gbw}) read: $\sigma_0 = 23.03\,\mb$,
$R_0(x) = 0.4\,\fm\times(x/x_0)^{0.144}$ with $x_0 = 3.04\times 10^{-4}$.

From Eq.~(\ref{UGD}) one can obtain explicitly the $\vec b$-dependent partial dipole-proton elastic amplitude performing integration in Eq.~(\ref{sat}),
%
\BA
\mathrm{Im} \mathcal{A}^N_{Q\bar Q}(\vec r, x, \alpha,\vec b\,)
=
\frac{\sigma_0}{8\pi \mathcal{B}(x)}\,
\Biggl\{
\exp\left[-\,\frac{\bigl [\vec b+\vec
r(1-\alpha)\bigr ]^2}{2\mathcal{B}(x)}\right] 
+ 
\exp\left[-\,\frac{(\vec
b-\vec r\alpha)^2}{2\mathcal{B}(x)}\right]
\nonumber\\
- \,2\,\exp\Biggl[-\,\frac{r^2}{R_0^2(x)}
-\,\frac{\bigl [\,\vec b+(1/2-\alpha)\vec
r\,\bigr ]^2}{2\mathcal{B}(x)}\Biggr]
\Biggr\}\,,
\nonumber\\
\label{dipa-gbw}
 \EA
%
where $\mathcal{B}(x)=R_N^2(x)+R_0^2(x)/8$.
Consequently, one can calculate the
forward ($t=0$) $t$-slope of the elastic dipole-proton cross section as, 
%
\BA
B_{Q\bar Q}(r,x)
= 
\frac{1}{2}\la b^2 \ra 
= 
\frac{1}{\sqq(r,x)}\int d^2b\,\,b^2\,
\mathrm{Im} \mathcal{A}^N_{Q\bar Q}(\vec r, x, \alpha,\vec b\,)\,.
\label{slope-qq} 
\EA
%

In Eq.~(\ref{dipa-gbw}) the function
$\mathcal{B}(x)$ can be simply determined from Eq.~(\ref{slope-qq}) 
assuming $\alpha = 1/2$.
Then we have,
%
\BA
B_{Q\bar Q}(r,x)
= 
\frac{1}{\sqq(r,x)}\int d^2b\,\,b^2\,
\mathrm{Im} \mathcal{A}^N_{Q\bar Q}(\vec r, x, \alpha=1/2,\vec b\,)
=
\mathcal{B}(x) + \frac{r^2}{8 \bigl [1- e^{-r^2/R_0^2(x)}\bigr ]}\,,
\label{bqq}
\EA
%
where the magnitude of $B_{Q\bar Q}(r,x)$ is probed at the well known scanning radius~ 
\cite{Nemchik:1996cw}
$r\approx r_S$
with $r_S = Y/\sqrt{Q^2 + m_V^2}$, and factors $Y$ have been determined in Ref.~\cite{Krelina:2018hmt}
for electroproduction of various heavy quarkonium states. 
However, treating electroproduction of heavy quarkonia, one can safely rely on the nonrelativistic limit of $\alpha = 1/2$ related to the scale parameter $Y\approx 6$ as was demonstrated in Ref.~\cite{Nemchik:1996cw}. Only in a case of charmonium electroproduction, small relativistic corrections at large $Q^2\gg m_V^2$ lead to a slow rise of $Y$ arriving at the value $Y\approx 7$ at $Q^2=100\,\GeV^2$. 

The magnitude of $B_{Q\bar Q}(r=r_S,x,Q^2)$ in Eq.~(\ref{bqq}) can be associated with the diffraction slope $B(\gamma^\ast\to V,x,Q^2)\equiv B_V(x,Q^2) = 
B_{Q\bar Q}(r_S,x,Q^2=0) - B_1\,\ln\,\bigl[ (m_{V}^2 + Q^2)/m_{\Jpsi}^2\bigr ]$, where $B_{Q\bar Q}(r_S,s,Q^2=0)$ with $s = (m_V^2 - t)/x$ conforms with the standard Regge form,
$B_{Q\bar Q}(r_S,s,Q^2=0)\equiv B_{\Jpsi}(s) = B_0 + 2\,\alpha^{\,\prime}(0)\,\ln(s/s_0)$. Here the parameters $B_0 = 2.0\,\GeV^{-2}$, $\alpha^{\,\prime} = 0.133\,\GeV^{-2}$ and $s_0 = 1\,\GeV^2$ have been obtained from the fit of the combined H1 \cite{Aktas:2005xu,Alexa:2013xxa} and ZEUS \cite{Breitweg:1997rg,Chekanov:2002xi} data. The value $B_1 = 0.45\,\GeV^{-2}$ has been determined in Ref.~\cite{Cepila:2019skb} analyzing electroproduction of $\Jpsi$ (see also Ref. \cite{jan-98}). 

The shape of the dipole amplitude (\ref{dipa-gbw}) correctly reproduces at $\vec q=0$ the dipole cross section according to Eq.~(\ref{dcs-b}). Moreover, such $\vec b$-dependent 
amplitude represents a source of an unique and additional information about details of the interaction mechanism since contains the color dipole orientation, which can be adapted for various phenomenological models for $\sqq(r,x)$ of a realistic saturated form given by Eq.~(\ref{gbw}). 
However,
there is a restriction of a simple GBW model related to an absence of the DGLAP evolution at large scales.
In order to eliminate this shortcoming,
besides the GBW dipole model, 
we take into account also  
the BGBK dipole model \cite{Bartels:2002cj} with the following modified parameter in Eq.~(\ref{dipa-gbw}),
%
\BA 
R_0^2(x)
\Rightarrow
R_0^2(x,\mu^2) = \frac{4}{Q_s^2(x,\mu^2)}
= \frac{\sigma_0\,N_c}{\pi^2\,\alpha_s(\mu^2)\,x\,g(x,\mu^2)},
\qquad\qquad
\mu^2 = \frac{\mathrel{C}}{r^2} + \mu_0^2\,,
\EA
%
where $Q_s^2$ is the saturation scale and the gluon distribution function $x\,g(x,\mu^2)$ is obtained as a solution of the DGLAP evolution equation
acquiring the subsequent parametrization at the initial scale $Q_0^2 = 1\,\GeV^2$,
%
\BA
x\,g(x,Q_0^2) 
= A_g \,x^{-\lambda_g} (1-x)^{5.6}\,.
\EA
%
Fitting procedure of the HERA data leads to the following model parameters:
%
\BA
A_g = 1.20,\qquad
\lambda_g = 0.28\,,\qquad
\mu_0^2 = 0.52\,\GeV^2\,\qquad
\mathrel{C} = 0.26\,\qquad
\sigma_0 = 23.0\,\mb\,.
\label{bgbk}
\EA
%
In what follows, the color dipole orientation, corresponding to $\vec b$-$\vec r$ correlation within the GBW and BGBK dipole model will be denoted in our paper as br-GBW and br-BGBK, respectively.

%
%
%
\section{Elastic electroproduction of quarkonia on protons}
\label{vm-proton}
%
%
%

Treating the elastic electroproduction of heavy quarkonia on proton targets, the corresponding differential cross section reads \cite{jan-98},
%
\BA
\frac{d\sigma^{\gamma^{\ast} p\to V p}(s,Q^2,t=-q^2)}{dt}
=
\frac{1}{16\,\pi}\,
\Bigl |
\mathcal{A}^{\gamma^{\ast} p\to V p}(s,Q^2, \vec q)
\Bigr |^2
=
\frac{1}{16\,\pi}\,
\Biggl |
\int\! d^2r\!\int\limits_0^1 \!d\alpha\,
\Psi_{V}^{*}(\vec r,\alpha)\,\!
\mathcal{A}_{Q\bar Q}(\vec r, s, \alpha,\vec q)\,\!
\Psi_{\gamma^{\ast}}(\vec r,\alpha,Q^2)
\Biggr |^2\,,
\nonumber\\
\label{proton}
\EA
%
where we take into account that
the partial dipole amplitude $\mathcal{A}_{Q\bar Q}$ can depend alternatively,
instead of the variable $x$ defined in Sect.~\ref{intro}, also on c.m. energy squared
$s = 2\,m_N\,p + m_N^2-Q^2$, where $m_N$ is the nucleon mass and $p$ is the photon energy in the rest frame
of the target. 

In Eq.~(\ref{proton}) the partial amplitude 
$\mathcal{A}_{Q\bar Q}(\vec r, s, \alpha,\vec q)$
is related to that given by Eq.~(\ref{dipa-gbw}) through the Fourier fransform,
%
\BA
\mathcal{A}_{Q\bar Q}(\vec r, s, \alpha,\vec q)
=
2\,\int d^2 b\,
e^{-i \vec{b}\cdot\vec{q}}\,\,
\mathrm{Im}\mathcal{A}^N_{Q\bar Q}(\vec r, s, \alpha,\vec b)\,
\EA
%
and correctly reproduces the dipole cross section $\sqq(r,s$) at $\vec q = 0$ as is given by Eq.~(\ref{dcs-b}).

Consequently, the final expression for the electroproduction amplitude 
$\mathcal{A}^{\gamma^{\ast} p\to V p}(s,Q^2, \vec q)$ in Eq.~(\ref{proton}) has the following
form,
%
\BA
\!\!\!\!\!\!
\mathcal{A}^{\gamma^{\ast} p\to V p}(s,Q^2, \vec q)
=
2\,\int d^2r\int\limits_0^1 d\alpha\,
\int d^2 b\,
\exp\Bigl [- i\,[\vec{b} + (1/2-\alpha)\,\vec{r}\,]\cdot\vec{q}\,\Bigr ]
\Psi_{V}^{*}(\vec r,\alpha)\,
\mathrm{Im} 
\mathcal{A}^N_{Q\bar Q}(\vec r, s, \alpha,\vec b)\,
\Psi_{\gamma^{\ast}}(\vec r,\alpha,Q^2)\,,
\label{amp-p}
\EA
%
where the argument in the exponential function takes into account the matter of fact that the transverse distance from
the centre of the proton to $Q$ or $\bar Q$ of the dipole is given by the distance $b$ to the center
of gravity of $Q\bar Q$ dipole and then by the relative distance of $Q$ or $\bar Q$ from the $Q\bar Q$-center of gravity. The latter distance varies with the fractional LF momenta as $r\alpha$ or $r(1-\alpha)$ and the corresponding $\alpha$-dependent part of the phase factor should vanish at $\alpha\to 1/2$.

Treating the electroprooduction of heavy quarkonia and assuming $s$-channel helicity conservation,
the $t$-dependent differential cross section (\ref{proton}) 
can be expressed as the sum of $T$ and $L$ contributions,
%
\BA
\frac{d\sigma^{\gamma^{\ast}p\rightarrow Vp}(s,Q^2,t=-q^2)}{dt}
&=&
\frac{d\sigma_T^{\gamma^{\ast}p\rightarrow Vp}(x,Q^2,t)}{dt} +
\tilde{\varepsilon}\,
\frac{d\sigma_L^{\gamma^{\ast}p\rightarrow Vp}(x,Q^2,t)}{dt}
\nonumber\\
&=&
\frac{1}{16\pi}\left(\Big\vert \mathcal{A}^{\gamma^{\ast}p\rightarrow Vp}_{T}(s,Q^2,\vec q~)\Big\vert^{2}+
\tilde{\varepsilon}\Big\vert \mathcal{A}^{\gamma^{\ast}p\rightarrow Vp}_{L}(x,Q^2,\vec q~)\Big\vert^{2}\right) \,,
\label{total-cs}
\EA
%
where we have taken the photon polarization $\tilde{\varepsilon} = 0.99$.

Our calculations include also a small real part 
\cite{bronzan-74,Nemchik:1996cw,forshaw-03} of the $\gamma^{\ast} p\to V p$ amplitude
performing the following replacement in Eq.~(\ref{amp-p}),
%
\BA
\mathcal{A}_{T,L}^{\gamma^{\ast} p\to V p}(s, \vec q)
\Rightarrow
\mathcal{A}_{T,L}^{\gamma^{\ast} p\to V p}(s, \vec q)
\,\cdot
\left(1 - i\,\frac{\pi\,\lambda_{T,L}}{2}\right)\,,
\qquad\qquad
\lambda_{T,L}
=
\frac
{\partial
 \,\ln\,{\mathcal{A}_{T,L}^{\gamma^{\ast} p\to V p}(s, \vec q)}}
{\partial\,\ln s}\ .
  \label{re/im}
\EA
%

Since the unintegrated off-diagonal gluon density $\mathcal{F}(x,\vec k,\vec k^{\,\prime})$ can contain gluons with different $x_2\ne x_1\approx x$ then, assuming a power-like behavior $\mathcal{F}(x)\propto x^{-\lambda_{T,L}}$ and small values of $x$, 
the modified skewed gluon distribution $\mathcal{\tilde{F}}(x_1,x_2\ll x_1,\vec k,\vec{k^{\,\prime}})$ is related to the conventional one as \cite{Shuvaev:1999ce},
%
\BA
\mathcal{\tilde{F}}(x_1,x_2\ll x_1,\vec k,\vec{k^{\,\prime}})
=
\mathcal{F}(x,\vec k,\vec{k^{\,\prime}})\,\cdot R_g(\lambda_{T,L})\,,
\EA
%
where the skewness factor $R_g$ reads,
%
\BA
R_g(\lambda_{T,L}) = 
\frac{2^{2\,\lambda_{T,L} + 3}}{\sqrt{\pi}} \,
\frac{\Gamma(\lambda_{T,L} + \frac{5}{2})}{\Gamma(\lambda_{T,L} + 4)}\,,
\label{skewness}
\EA
%
and functions $\lambda_{T,L}$ are given by Eq.~(\ref{re/im}). Such the skewness effect is included in our
calculations via substitution $\mathcal{F}(x,\vec k,\vec{k^{\,\prime}})\Rightarrow \mathcal{F}(x,\vec k,\vec{k^{\,\prime}})\cdot R_g(\lambda_{T,L})$.

For a simple ``$S$-wave-only'' structure of the $V\to Q\bar Q$ vertex in the $Q\bar Q$ rest frame
\cite{Ivanov:2002kc,Ivanov:2002eq,Ivanov:2007ms,Krelina:2018hmt,Cepila:2019skb} the spin-dependent component of the quarkonium wave function undergoes 
the Melosh spin transformation to the LF frame,
what leads to the following specific form of electroproduction amplitudes given by Eq.~(\ref{amp-p}) for $T$ and $L$ polarizations,
%
\BA
\mathcal{A}_T^{\gamma^{\ast} p\to V p}(s,Q^2,\vec{q})
&=&
N_p\,
\int d^2r
\int_0^1 d\alpha  \,
\int d^2 b\,
\exp\Bigl [- i\,[\vec{b} + (1/2-\alpha)\,\vec{r}\,]\cdot\vec{q}\,\Bigr ]\,
\nonumber\\
&\times&
\mathrm{Im} 
\mathcal{A}^N_{Q\bar Q}(\vec r, s, \alpha,\vec b)\,
\left[\Sigma^{(1)}_{T}(r,\alpha,Q^2) 
+
\Sigma^{(2)}_{T}(r,\alpha,Q^2)\right]\,,
 \nonumber
\\ 
\!\!\!\!\!\!\!\!\!\!
   \Sigma^{(1)}_T(r,\alpha,Q^2)
   &=&  
   K_0(\varepsilon r) \int_0^\infty dp_T\,p_T\,
   J_0(p_T r) \Psi_V (\alpha,p_T) 
   \left[ \frac{2\,m_Q^2(m_L+m_T)+m_L\,p_T^2}{ m_T (m_L + m_T)} \right]\,,
\nonumber \\
\!\!\!\!\!\!\!\!\!\!
  \Sigma^{(2)}_T(r,\alpha,Q^2)
  &=& 
   K_1(\varepsilon r) \int_0^\infty dp_T\,p_T^2\,
   J_1(p_T r) \Psi_{V} (\alpha,p_T) \left[
\,\varepsilon\, 
   \frac{m_Q^2(m_L+2m_T)-m_T\,m_L^2}{m_Q^2\,m_T (m_L+m_T)} \right]\,,
   \label{amp-p-sr1}
\EA
%
and 
%
\BA
\mathcal{A}_L^{\gamma^{\ast} p\to V p}(s,Q^2,\vec{q})
&=&
N_p\,
\int d^2r
\int_0^1 d\alpha  \,
\int d^2 b\,
\exp\Bigl [- i\,[\vec{b} + (1/2-\alpha)\,\vec{r}\,]\cdot\vec{q}\,\Bigr ]\,
\mathrm{Im} 
\mathcal{A}^N_{Q\bar Q}(\vec r, s, \alpha,\vec b)\,
\Sigma_{L}(r,\alpha,Q^2)\,,
 \nonumber
\\ 
\!\!\!\!\!\!\!\!\!\!
   \Sigma_L(r,\alpha,Q^2)
   &=&  
   K_0(\varepsilon r) \int_0^\infty dp_T\,p_T\,
   J_0(p_T r) \Psi_V (\alpha,p_T) 
   \left[ 4\,Q\,\alpha\,(1-\alpha)\,\frac{m_Q^2 + m_L m_T}{m_Q\,(m_L+m_T)}
   \right]\,,
   \label{amp-p-sr2}
\EA
%
where $N_p=Z_Q\,\sqrt{2 N_c \,\alpha_{em}}/2\,\pi$,
$\varepsilon^2 = m_Q^2\,+ \alpha (1-\alpha) Q^2$,
$\aem= 1/137$ is the fine-structure constant, the factor $N_c=3$ represents the number of colors in QCD, $Z_Q=2/3$ and $1/3$ are the charge isospin factors for charmonium and bottomonium production, respectively,  
variables
$m_T = \sqrt{m_Q^2 + p_T^2}$
and 
$m_L = 2\, m_Q\,\sqrt{\alpha(1-\alpha)}$\,,
and
$J_{0,1}$ and $K_{0,1}$ are the Bessel functions of the first kind and the modified Bessel functions of the second kind, respectively.

The generalization of the color dipole factorization formula (\ref{amp-p}) with Eq.~(\ref{amp-p-sr1}) for $T$ polarised and with Eq.~(\ref{amp-p-sr2}) for $L$ polarised photons to the diffraction slope of the process $\gamma^{\ast}~p\to V~p$ is straightforward and reads,
%
\BA
B_{T,L}(\gamma^{\ast}\to V,s,Q^2)\,\cdot\,
\mathcal{A}_{T,L}^{\gamma^{\ast} p\to V p}(s,Q^2,\vec{q}=0)
&=& 
\int d^2r\int_0^1 d\alpha\,
B_{Q\bar Q}(r,s)\,\sqq(r,s)\,
\Psi_{V}^{*}(\vec r,\alpha)_{T,L}\,
\Psi_{\gamma^{\ast}}(\vec r,\alpha,Q^2)_{T,L}\,,
\nonumber\\
B_T(\gamma^{\ast}\to V,s,Q^2)\,\cdot\,
\mathcal{A}_T^{\gamma^{\ast} p\to V p}(s,Q^2,\vec{q}=0)
&=&
\frac{N_p}{2}\,
\int d^2r\int_0^1 d\alpha\,\int d^2 b\,b^2\,
\mathrm{Im}\mathcal{A}^N_{Q\bar Q}(\vec r, s, \alpha,\vec b)\,
\nonumber\\
&\times&
\left[\Sigma^{(1)}_{T}(r,\alpha,Q^2) 
+
\Sigma^{(2)}_{T}(r,\alpha,Q^2)\right]\,,
\nonumber\\
B_L(\gamma^{\ast}\to V,s,Q^2)\,\cdot\,
\mathcal{A}_L^{\gamma^{\ast} p\to V p}(s,Q^2,\vec{q}=0)
&=&
\frac{N_p}{2}\,
\int d^2r\int_0^1 d\alpha\,\int d^2 b\,b^2\,
\mathrm{Im}\mathcal{A}^N_{Q\bar Q}(\vec r, s, \alpha,\vec b)\,
\Sigma_{L}(r,\alpha,Q^2) 
\,,
\label{slope-vm}
\EA
%
where the forward dipole $t$-slope $B_{Q\bar Q}(r,s)$ is given by Eq.~(\ref{slope-qq}).
Consequently, in comparison with the method described in the previous section,
Eq.~(\ref{slope-vm}) represents more exact way for determination of functions $\mathcal{B}(x)$ in elastic partial dipole amplitude Eq.~(\ref{dipa-gbw}). However, corresponding differences in acquiring of $\mathcal{B}(x)$ are very small, $\lsim 1\%-2\%$, and we rely on a more simple way based on the scanning radius phenomenon (see Sec.~\ref{sat-cdm}).
Besides,
here we have found that differences $B_T-B_L$, for given values of $\mathcal{B}(x)$ extracted by a more simple method, are tiny as well and do not exceed $0.01$ and $0.1\,\GeV^{-2}$ for electroproduction of 1S and 2S charmonium states, respectively, what is in accordance with results from Ref.~\cite{jan-98}.

Finally, we would like to emphasize that the Lorentz boost of radial components of quarkonium wave functions to the LF frame has been based on the Terent’ev prescription \cite{Terentev:1976jk}, which represents only ad hoc procedure for such a transformation.
However, its validity has been proven in Ref.~\cite{Kopeliovich:2015qna}
by comparing with calculations employing the Lorentz-boosted Schr\"odinger equation, which was found rather accurate for heavy quarks

%
%
%
\section{Model predictions vs. data}
\label{data-proton}
%
%
%

In this section, our model calculations for differential cross sections $d\sigma/dt$ are tested by comparing with available data on electroproduction of $\Jpsi(1S)$ from experiments at HERA.
We present also model predictions for another quarkonium states, such as $\psip(2S)$, $\Y(1S)$ and $\Yp(2S)$. 
Although quarkonium wave functions can be generated by several distinct models for the $Q$-$\bar Q$ interaction potential (see Refs.~\cite{Buchmuller:1980su} for the Buchm\"uller-Tye potential ({BT}), \cite{Barik:1980ai} for the effective power-law potential ({Pow}),
\cite{Quigg:1977dd} for the logarithmic potential ({Log}), \cite{Eichten:1978tg,Eichten:1979ms} for the Cornell potential ({Cor}) and \cite{Cepila:2019skb} for the 
harmonic oscillatory potential ({HO})), in our analysis we adopt only two of them, BT and Pow, since they provide the best description of available data as was shown in Ref.~\cite{Cepila:2019skb}.

In calculations we include two sets of models. The first group is related to
br-GBW and br-BGBK models (see Sec.~\ref{sat-cdm}) with $\vec b$-dependent partial dipole amplitude based on realistic color dipole orientation and leading to a proper correlation between vectors $\vec b$ and $\vec r$. The second class of models, widely used in the literature, contains $b$-dependent part, which is implemented additionally into conventional phenomenological models for the dipole cross section, like b-IPsat model \cite{Rezaeian:2012ji}, b-Sat model \cite{Kowalski:2003hm}, b-CGC model
\cite{Kowalski:2006hc,Rezaeian:2013tka},
and/or such $\vec b$-$\vec r$ correlation is not included properly, like in the b-BK model from Ref.~\cite{Cepila:2018faq} where
an angle between these vectors is set to zero.

\BF
\PSfig{0.865}{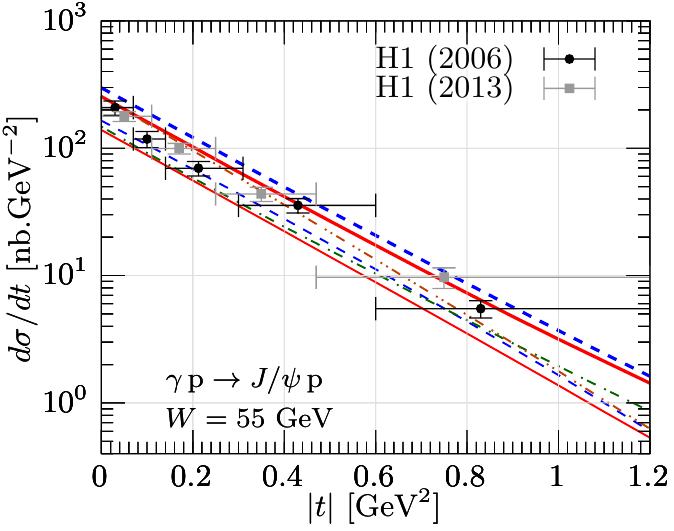}~~~
\hspace*{-0.5cm}
\PSfig{0.865}{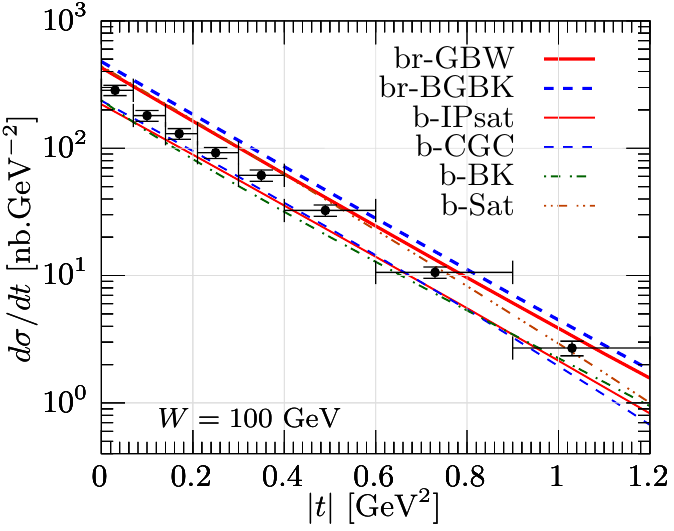}~~~~
\hspace*{-0.55cm}
\PSfig{0.865}{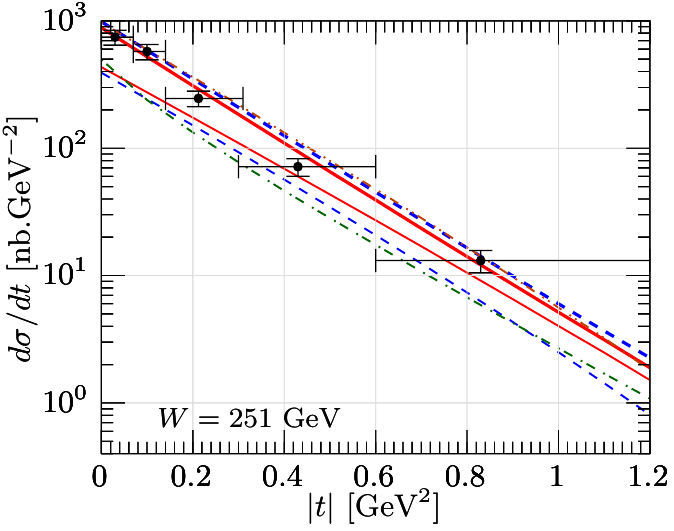}~~~~
%
\Caption{
  \label{Fig-tdep-data1}
  Comparison of our predictions for the momentum transfer dependence of the differential cross sections $d\sigma^{\gamma p\to \Jpsi p}(t)/dt$ with HERA data from the
  H1 \cite{Aktas:2005xu,Alexa:2013xxa} collaboration
  at different c.m. energies $W=55, 100$ and $251\,\GeV$. Different curves correspond to our results using
  br-GBW and br-BGBK models (thick lines) for $\vec b$-dependent partial dipole amplitude including a proper $\vec b$-$\vec r$ correlation (see Eq.~(\ref{dipa-gbw})), as well as b-IPsat, b-CGC, b-Sat and b-BK models (thin lines) 
  with only additional factorized $b$-dependence and with the approximation $\vec b\cdot\vec r = b r$, respectively. 
  The quarkonium wave functions are generated by the BT potential.
  }
\EF

Figure~\ref{Fig-tdep-data1} shows model predictions for differential cross sections $d\sigma^{\gamma p\to \Jpsi p}/dt$ in comparison with available HERA data from 
the H1 \cite{Aktas:2005xu,Alexa:2013xxa} collaboration at several fixed values of c.m. energy $W = 55, 100$ and $251\,\GeV$. 
Here the charmonium wave function is determined from the BT potential.
Besides,
our results based on the br-GBW and br-BGBK dipole models with a proper $\vec b$-$\vec r$ correlation are tested by comparing with several conventional $b$-dependent dipole models where such correlation is not included accurately (denoted as b-IPsat, b-CGC, b-BK and b-Sat). 

One can see from Fig.~\ref{Fig-tdep-data1} 
that dipole models describing differently the data can be divided into two groups corresponding to br-GBW, br-BGBK and b-Sat vs b-IPsat, b-CGC and b-BK models. The models from the former group exhibit a little bit better
agreement with H1 data through the all energy region from $W=55\,\GeV$ to $251\,\GeV$.
Especially at higher $W=251\,\GeV$ the models without a proper $\vec b$-$\vec r$ correlation substantially underestimate the data. 

\BF
\PSfig{0.88}{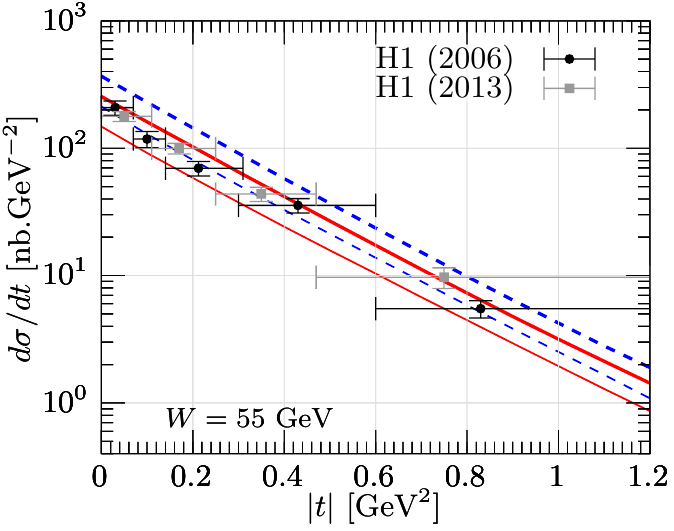}~~~~
\hspace*{-0.75cm}
\PSfig{0.88}{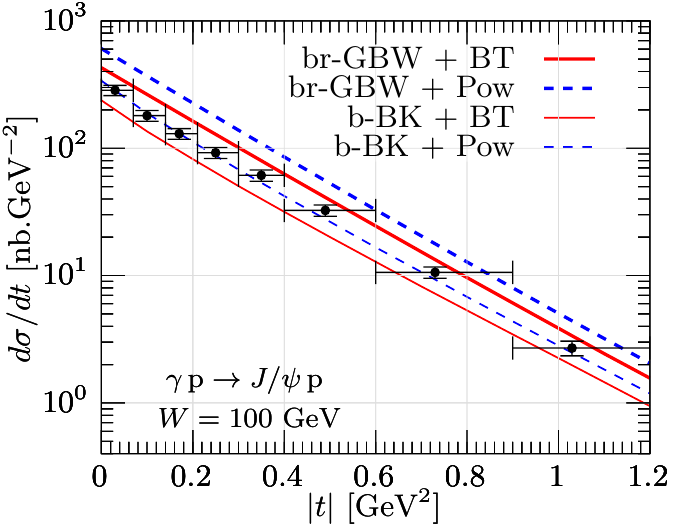}~~~~
\hspace*{-0.75cm}
\PSfig{0.88}{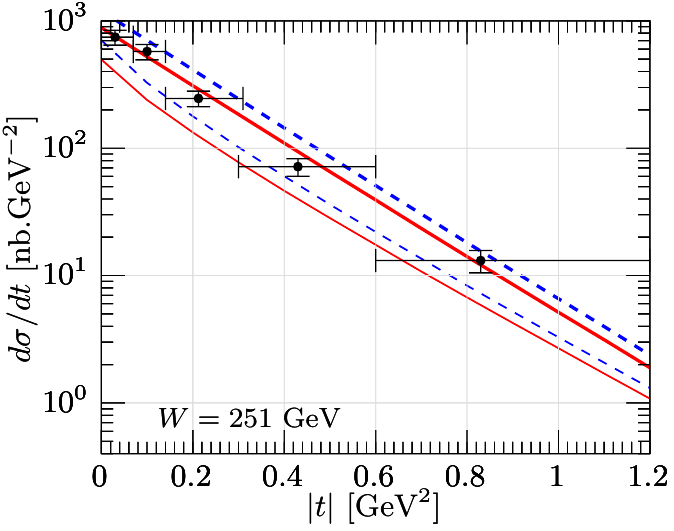}~~~~\\
%
\Caption{
  \label{Fig-tdep-data2}
  The same as Fig.~\ref{Fig-tdep-data1} but for the $b$-dependent partial dipole amplitude generated by br-GBW (thick lines) and b-BK (thin lines) dipole models and for quarkonium wave functions determined from the BT (solid lines) and Pow (dashed lines) $c$-$\bar c$ interquark potentials. Our predictions are compared with H1 data \cite{Aktas:2005xu,Alexa:2013xxa}
  at c.m. energy $W = 55\,\GeV$ (left panel), $100\,\GeV$ (middle panel) and $251\,\GeV$ (right panel).
  }
\EF

Besides of a sensitivity of predictions to various $b$-dependent dipole models analyzed in Fig.~\ref{Fig-tdep-data1}, we demonstrate that
the magnitude of  $d\sigma^{\gamma p\to \Jpsi p}/dt$ is strongly correlated with the shape of $c$-$\bar c$ interaction potential. As an example, we show in Fig.~\ref{Fig-tdep-data2} our results for the BT (solid lines) and Pow (dashed lines)  potentials. One can see that calculations with the charmonium wave function generated by the BT model exhibit a better description of H1 data at larger $W$.
Since both models for partial dipole amplitude, br-GBW and br-BGBK, give similar results for $d\sigma/dt$ (see Fig.~\ref{Fig-tdep-data1}), here we confront our predictions using only br-GBW (thick lines) with results based on a popular b-BK (thin lines) dipole model at different c.m. energies $W=55, 100$ and $251\,\GeV$. Treating the BT potential, in comparison with the former model, the calculations based on b-BK dipole amplitude show evidently a worse description of data exhibiting their underestimation, which is stronger at higher photon energies.

\BF
\PSfig{0.87}{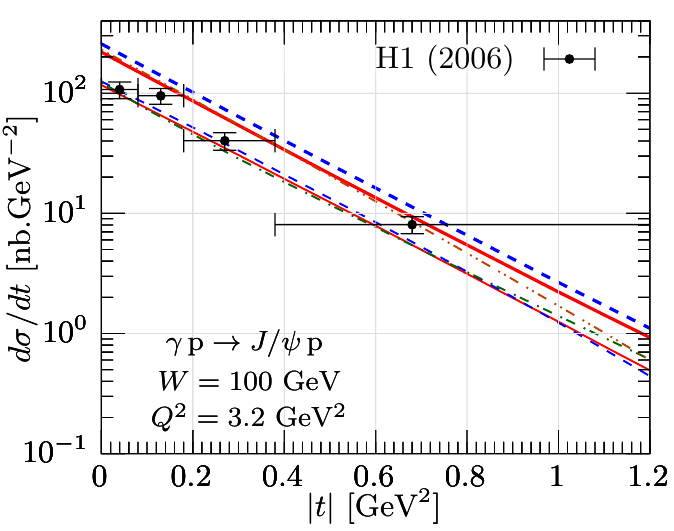}~~~~
\hspace*{-0.75cm}
\PSfig{0.87}{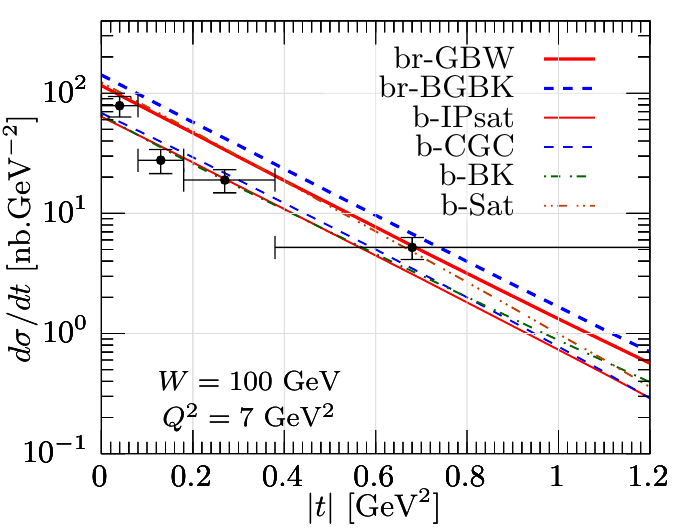}~~~~
\hspace*{-0.75cm}
\PSfig{0.87}{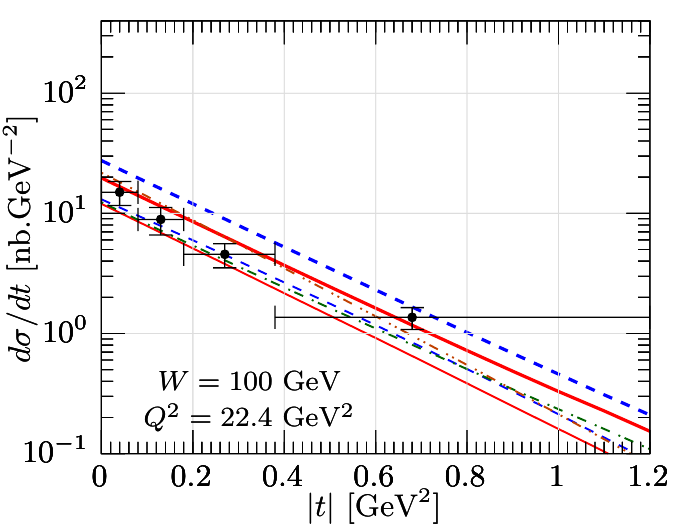}~~~~\\
\Caption{
  \label{Fig-tdep-data1-s}
  Comparison of our predictions for differential cross sections $d\sigma^{\gamma^\ast p\to \Jpsi p}(t)/dt$ with H1 data \cite{Aktas:2005xu} at different photon virtualities $Q^2=3.2\,\GeV^2$ (left panels), $Q^2 = 7.0\,\GeV^2$ (middle panels) and $Q^2 = 22.4\,\GeV^2$ (right panels). Different lines correspond to our results at $W = 100\,\GeV$ using various models for $b$-dependent partial dipole amplitudes and the quarkonium wave function generated by the BT potential. 
  }
\EF

Except for the real photoproduction process ($Q^2=0$),
another test for $b$-dependent dipole models
concerns to virtual electroproduction of charmonia with $Q^2 > 0$. The corresponding model predictions are compared with H1 data \cite{Aktas:2005xu} in Fig.~\ref{Fig-tdep-data1-s} at c.m. energy $W=100\,\GeV$ and at several photon virtualities $Q^2 = 3.2\,\GeV^2$ (left panels), $Q^2=7.0\,\GeV^2$ (middle panels) and $Q^2=22.4\,\GeV^2$ (right panels).
Here all panels show our results adopting different dipole models and taking the charmonium wave function generated by the BT potential.
From Fig.~\ref{Fig-tdep-data1-s} one cannot give any definite conclusion which models provide the best description of H1 data through the all region of $Q^2$. The new more precise data from future electron-proton colliders can help us to rule out various models for $b$-dependent partial dipole amplitudes, as well as for charmonium wave functions.

\BF
\PSfig{0.90}{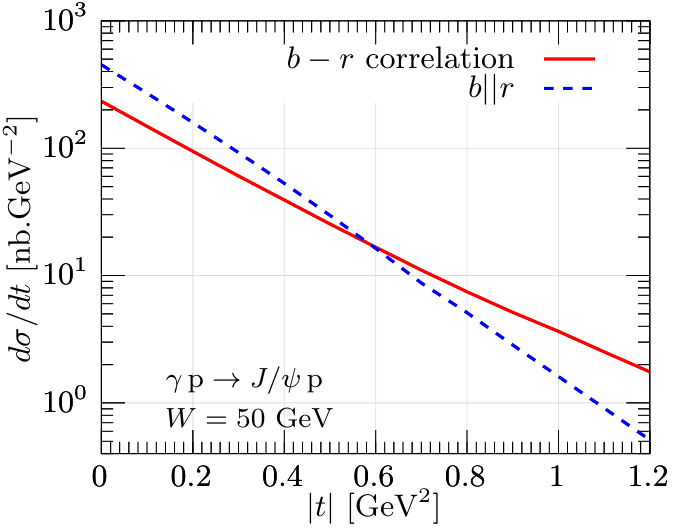}
\hspace*{0.3cm}
\PSfig{0.90}{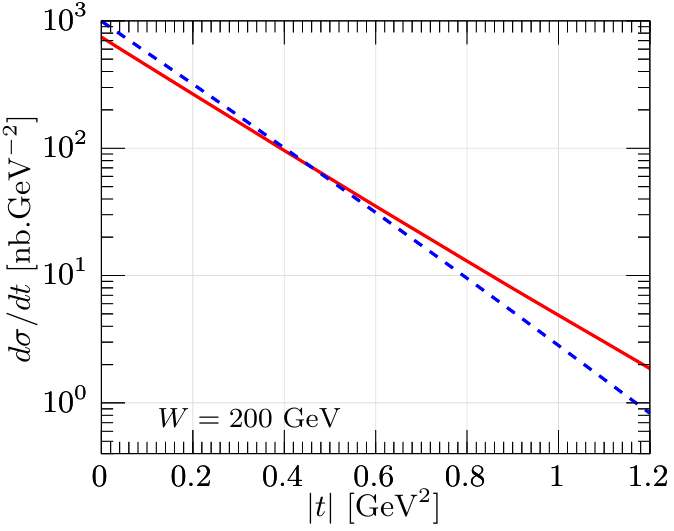}\\
\PSfig{0.90}{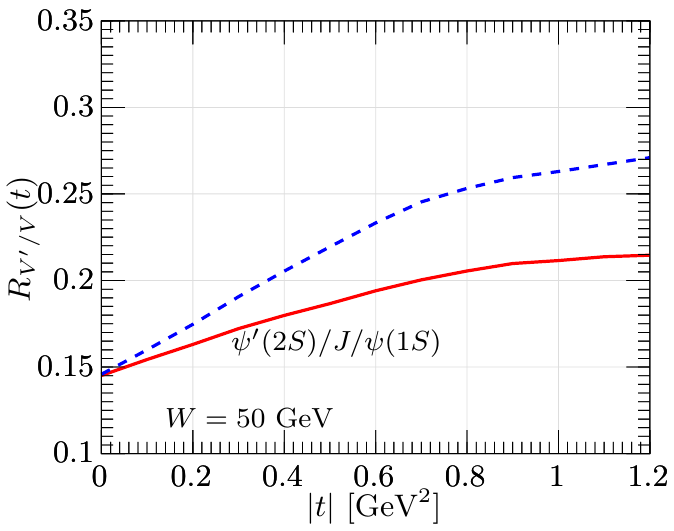}
\hspace*{0.3cm}
\PSfig{0.90}{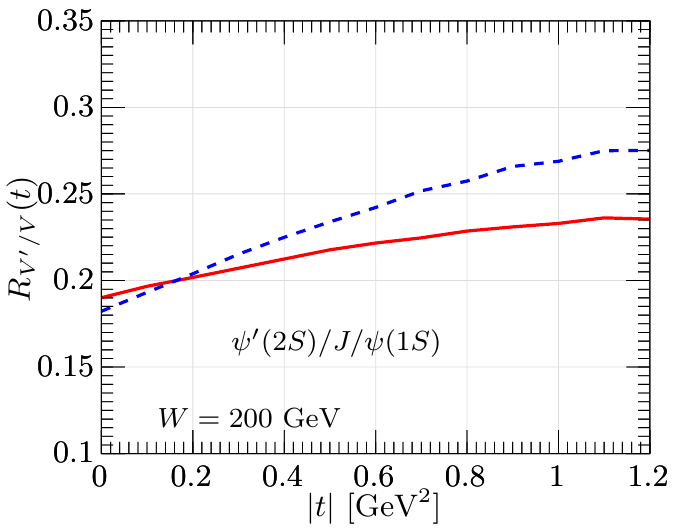}
\Caption{
  \label{Fig-tdep-corr}
  Demonstration of importance of the $\vec b$-$\vec r$ correlation in the partial elastic dipole amplitude (\ref{dipa-gbw}) by performing calculations of differential cross sections
  $d\sigma^{\gamma p\to\Jpsi p}(t)/dt$ (top panels) and $\psip(2S)$-to-$\Jpsi(1S)$ ratio of the differential cross sections 
  $R_{V'(2S)/V(1S)}(W,t) = \{d\sigma^{\gamma p\to V'(2S) p}/dt\}/ \{d\sigma^{\gamma p\to V(1S) p}/dt\}$
  (bottom panels) at c.m. energy $W=50\,\GeV$ (left panels) 
  and $200\,\GeV$ (right panels). The corresponding results based on the br-GBW dipole model (solid lines) are compared with the case when vectors $\vec b$ and $\vec r$ are parallel (dashed lines). The charmonium wave function is determined from the BT potential. 
  }
\EF

In the next Fig.~\ref{Fig-tdep-corr} we demonstrate how the model results are modified at $W=50$ (left panel)
and $200\,\GeV$ (right panel) taking a realistic $\vec b$-$\vec r$ correlation in the partial dipole amplitude (\ref{dipa-gbw}) (solid lines) in comparison with a 
simplified
assumption when $\vec b$$\parallel$$\vec r$ (dashed lines). Our calculations of $d\sigma^{\gamma p\to\Jpsi(1S) p}(t)/dt$ (top panels) and $\psip(2S)$-to-$\Jpsi(1S)$ ratio of differential cross sections (bottom panels) have been performed with the br-GBW dipole model and charmonium wave function generated by the BT potential. One can see that incorporation of a proper $\vec b$-$\vec r$ correlation leads to a sizeable modification of the corresponding $t$-dependence, what has an indispensable impact on all predictions based on the b-BK model \cite{Cepila:2018faq} where authors assume that the dipole amplitude is independent of the angle between vectors $\vec b$ and $\vec r$. Consistently with Fig.~\ref{Fig-tdep-corr}, if predictions incorporating this b-BK dipole model have provided a good description of data on diffractive electroproduction of vector mesons on protons and nuclei, the subsequent incorporation of a realistic correlation between vectors $\vec b$ and $\vec r$ will spoil such a good agreement with data. In another words, the corresponding $t$-slope of $d\sigma^{\gamma^\ast p(A)\to V p(A)}(t)/dt$ becomes smaller keeping the same model parameters.
The Fig.~\ref{Fig-tdep-corr} also demonstrates that the onset of a proper $\vec b$-$\vec r$ correlation with respect to a simplified $\vec b$$\parallel$$\vec r$-case becomes stronger towards smaller photon energies.

\BF
\hspace*{0.4cm}
\PSfig{0.90}{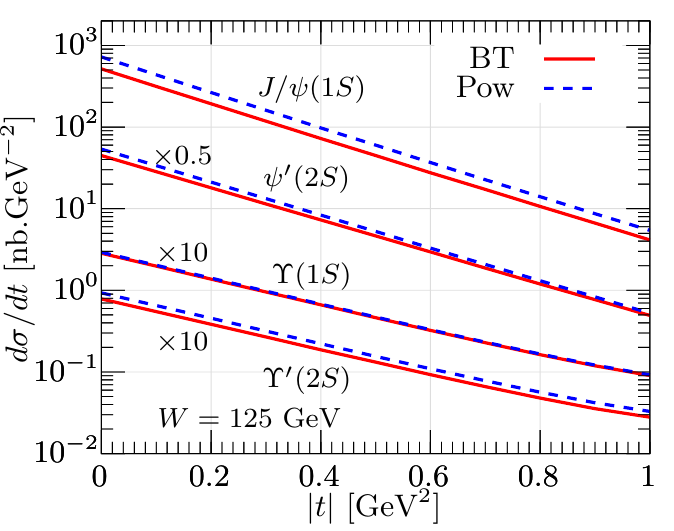}~~~~
\PSfig{0.90}{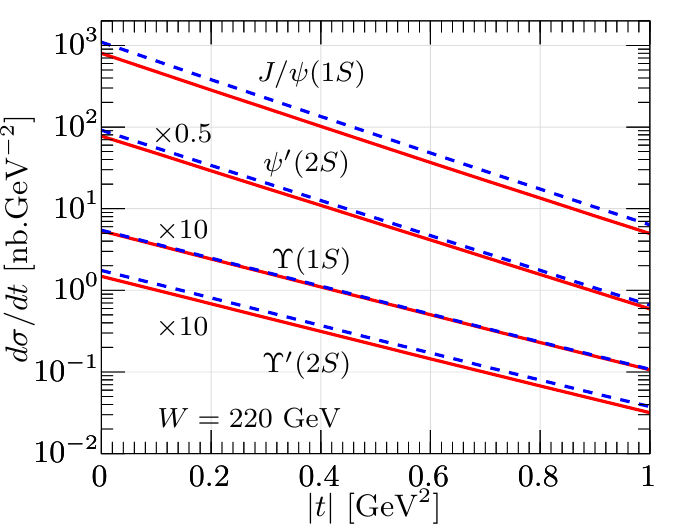}~~~~
%
\Caption{
  \label{Fig-tdep-data3}
  The model predictions for $t$-dependent differential cross sections of photoproduction of different quarkonium states at c.m. energy $W = 125$ (left panel) and $220\,\GeV$ (right panel). Our calculations have been performed adopting the br-GBW model for the partial dipole amplitude 
  taking the BT (solid lines) and Pow (dashed lines) models for
  $c$-$\bar c$ and $b$-$\bar b$ interaction potentials.   
  }
\EF

Figure~\ref{Fig-tdep-data3} shows our results of $d\sigma/dt$ for production of various quarkonium states, such as $\Jpsi(1S)$, $\psip(2S)$, $\Y(1S)$ and $\Yp(2S)$. The corresponding predictions are depicted at two c.m. energies $W=125$ and $220\,\GeV$ that can be scanned by recent experiments at the LHC, as well as by the future measurements at electron-proton colliders. 
Here we present also a sensitivity of calculations to quarkonium wave functions generated by BT and Pow models for $Q$-$\bar Q$ interaction potentials.  One can see that corresponding theoretical uncertainties
are reduced in production of bottomonium states due to a smaller variance in determination of the $b$-quark mass by the BT and Pow $b$-$\bar b$ potential models used in our analysis. 
Differences between solid and dashed lines related to various quarkonium wave functions can be treated as a measure of the theoretical uncertainty.

\BF
\PSfig{0.87}{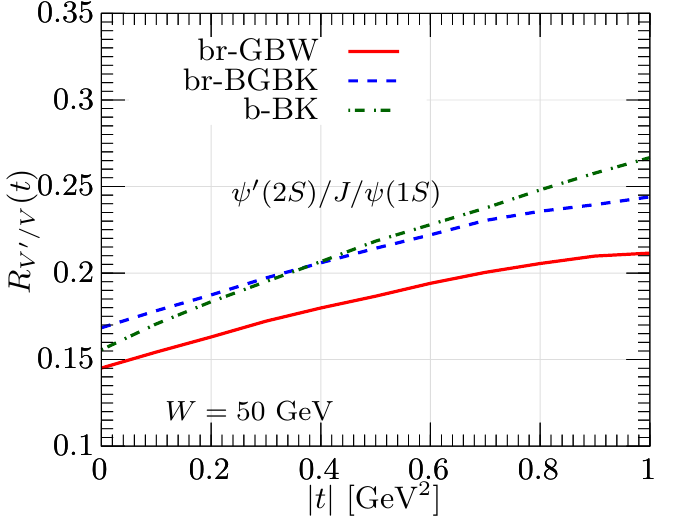}~~~~
\hspace{-0.6cm}
\PSfig{0.87}{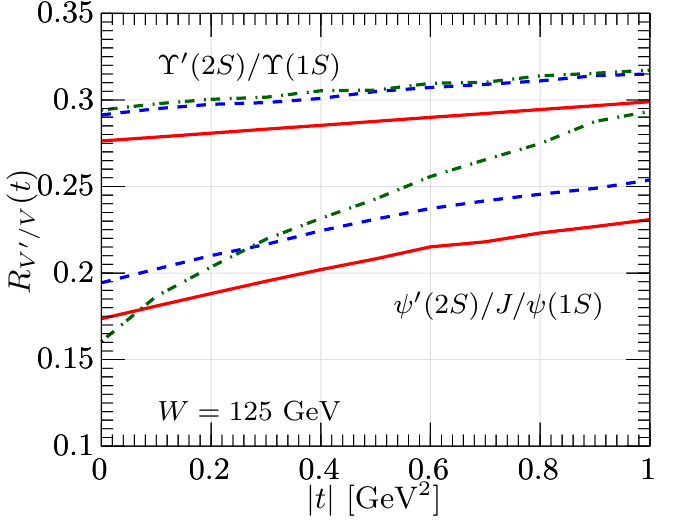}~~~~
\hspace*{-0.6cm}
\PSfig{0.87}{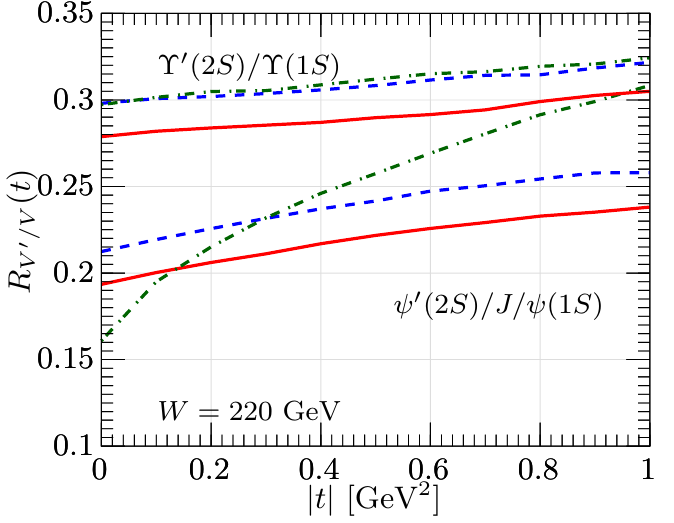}~~~~
\Caption{
  \label{Fig-tdep-data4}
    The model predictions for the $t$-dependent $V'(2S)$-to-$V(1S)$ ratio of differential cross sections
  $R_{V'/V}(W,t)$ at $Q^2=0$. Ratios $\psip(2S)/\Jpsi(1S)$ and $\Y'(2S)/\Y(1S)$ are depicted at several c.m. energies $W=50, 125, 220\,\GeV$ and $W=125, 220\,\GeV$, respectively. The quarkonium wave functions are generated by the BT potential. The solid, dashed and dot-dashed lines correspond to $b$-dependent partial dipole amplitude obtained from br-GBW, br-BGBK and b-BK dipole models, respectively. 
  }
\EF

The node effect is demonstrated in charmonium production as an inequality $B(\psip(2S)) < B(\Jpsi(1S))$ and its manifestation can be studied in terms of the $t$-dependent differential cross section ratio $R_{V'(2S)/V(1S)}(W,t) = \{d\sigma^{\gamma p\to V'(2S) p}/dt\}/ \{d\sigma^{\gamma p\to V(1S) p}/dt\}$
for real photoproduction. One can see from Fig.~\ref{Fig-tdep-data4} that, as a consequence of the node effect, the ratio $R_{\psip/\Jpsi}(t)$ rises with $t$ at $W= 50\,\GeV$. However, at higher $W\gsim 100\,\GeV$ this rise is changed gradually for a 
more flat 
$t$-behavior of $R_{\psip/\Jpsi}(t)$ and $R_{\Yp/\Y}(t)$ as a result of a weaker node effect at larger energies and for heavier vector mesons, respectively. So such expected scenario is confirmed by our results based on br-GBW and br-BGBK models and is in correspondence with analysis from Ref.~\cite{jan-98}.

Figure~\ref{Fig-tdep-data4} also nicely confirms that the study of $t$-dependent $\psip(2S)/\Jpsi(1S)$ ratio represents a very effective tool for ruling out various $\vec b$-dependent models for the partial elastic dipole amplitude, especially if $\vec b$-$\vec r$ correlation is not included properly. As an example, we discuss here a popular b-BK model where the dipole amplitude is acquired for the case $\vec b$$\parallel$$\vec r$ \cite{Cepila:2018faq}. The corresponding predictions are plotted by dot-dashed lines. One can see that the rise of $R_{\psip/\Jpsi}(t)$ is stronger at larger $W$ and is much more intensive
in comparison with the flat $t$-behavior obtained within br-GBW and br-BGBK models. Besides, the ratio of forward cross sections $R_{\psip/\Jpsi}(t=0)$ practically does not depend on energy $W$.
Such results are unexpected, they are in contradiction with our expectations and cannot be proven by any physical reasons. In another words, they correspond to a rise with energy $W$ of a difference between slope parameters $B(\Jpsi(1S)) - B(\psip(2S))$, what is not conformed with any physical interpretation. 
The correct partial explanation of this puzzle is based on an absence of a proper correlation between vectors $\vec b$ and $\vec r$ in calculations based on the b-BK model \cite{Cepila:2018faq}. This means that all related predictions for $d\sigma/dt$ adopting this model with condition $\vec b$$\parallel$$\vec r$ are not accurate. In order to demonstrate this conclusion we have presented in bottom panels of Fig.~\ref{Fig-tdep-corr}, as an example, also calculations of $R_{\psip/\Jpsi}(t)$ obtained from our br-GBW model treating also the case $\vec b\cdot\vec r = br$ like in the b-BK model.
One can see that 
corresponding results have been changed significantly exhibiting now much stronger rise of the ratio $R_{\psip/\Jpsi}(t)$ with $t$ which is more pronounced at smaller energies. Thus, such an effect allows to conclude that the investigation of $t$-dependent behavior of $R_{\psip/\Jpsi}(t)$ at $Q^2=0$ is very suitable for an analysis of a correlation between $\vec b$ and $\vec r$ since in this case the larger dipole sizes of 2S states generate more sensitive correlations with the impact parameter of a collision.

Figure~\ref{Fig-tdep-data4} also demonstrates that rather large sensitivity of model predictions for $R_{\psip/\Jpsi}(t)$ to $\vec b$-$\vec r$ correlation is melt away in the bottomonium case due to much smaller dipole sizes $r(\Yp(2S))\ll r(\psip(2S))$ contributing to the corresponding diffraction process. For this reason the difference between calculations with a proper $\vec b$-$\vec r$ correlations (br-GBW and br-BGBK models) and results based on a simplified assumption $\vec b$$\parallel$$\vec r$ (b-BK model) is much smaller. Specifically, results with the br-BGBK dipole amplitude almost coincide with values from the b-BK model. Besides, both types of predictions for $R_{\Yp/\Y}(t)$ exhibit a similar $t$-shape, which is also in accordance with expected more flat $t$-dependence at higher photon energies and for heavier quarkonia as a manifestation of a weaker node effect.

%
%
%
\section{Conclusions}
\label{final}
%
%
%

We study the momentum transfer dependence of the differential cross section for diffractive electroproduction of heavy quarkonia on protons. Our main results are as follows:

\begin{itemize}

    \item 
Basing on our the previously developed models for the $b$-dependent partial
elastic dipole-proton amplitude including the $\vec b$-$\vec r$ correlation, we calculated the $t$-dependent differential cross sections of diffractive production of various quarkonium states. The results are confronted with the widely used phenomenological models, which either miss the $\vec b$-$\vec r$ correlation or do not incorporate it properly. 
    
    \item
The radial component of the quarkonium wave function was generated in the $Q\bar Q$ rest frame by solving the Schr\"odinger equation with various popular models for the $Q$-$\bar Q$ potential. The LF counterpart is then obtained applying a Lorentz boost procedure, which validity for heavy dipoles was confirmed in \cite{Kopeliovich:2015qna}. Here we also included the Melosh effect of spin rotation, which significantly affects the production cross section.

    \item
Our model predictions for $d\sigma^{\gamma p\to \Jpsi p}/dt$ were successfully tested  comparing  with available data from the H1 
experiment at HERA  at different c.m. energies $W$ (Fig.~\ref{Fig-tdep-data1}) and photon virtualities $Q^2$ (Fig.~\ref{Fig-tdep-data1-s}). The models, labelled as br-GBW and br-BGBK, based on the realistic $\vec b$-$\vec r$ correlation, exhibit a better description of HERA data, compared with the conventional $b$-dependent dipole models, like b-IPsat, b-CGC, b-BK and b-Sat (see also Fig.~\ref{Fig-tdep-data2}), which do not include properly such a correlation. Specifically, the calculations performed with the popular b-BK model, employ a strongly exaggerated  strength of the $\vec b$-$\vec r$ correlation \cite{Cepila:2018faq} what significantly  affects the differential cross section, especially for radially excited charmonium states $\psip(2S)$ (Fig.~\ref{Fig-tdep-data4}). In particular, it leads to a larger $t$-slope of $d\sigma^{\gamma p\to\Jpsi p}/dt$, and predicts a much stronger rise of the $\psi(2S)/\Jpsi$ ratio with $t$, especially at smaller photon energies, as was demonstrated in Fig.~\ref{Fig-tdep-corr}.

    \item
We predicted the differential cross sections of photoproduction of various quarkonium states (Fig.~\ref{Fig-tdep-data3}) that can be verified in UPC collisions at the LHC, as well as with future experiments at electron-proton colliders.

    \item
All expressions for $t$-dependent differential cross sections for quarkonium electroproduction on protons  can be generalized for nuclear targets. The corresponding predictions  for the forthcoming UPC measurements at the LHC, and future electron-ion colliders,
will be presented elsewhere \cite{prepar}.

    \item
The effect of $\vec b$-$\vec r$ correlation leads to a specific polarization of the produced quarkonia, which can be observed in the polar angle asymmetry of the dileptons from  quarkonium decays. This effect will be studied in a separate paper.

\end{itemize}

\noindent\textbf{Acknowledgment}: 
This work was supported in part by grants ANID - Chile FONDECYT 1170319 and ANID PIA/APOYO AFB180002.
The work of J.N. was partially supported by Grant
No. LTT18002 of the Ministry of Education, Youth and
Sports of the Czech Republic,
by the project of the
European Regional Development Fund No. CZ.02.1.01/0.0/0.0/16\_019/0000778
and by the Slovak Funding Agency, Grant No. 2/0007/18.
The work of M.K. was supported in part 
by the project of the
European Regional Development Fund No. CZ.02.1.01/0.0/0.0/16\_019/0000778
and by the International Mobility of Researchers - MSCA IF IV at CTU in Prague 
CZ.02.2.69/0.0/0.0/20\_079/0017983, Czech Republic.



\end{document}